\documentclass[12pt]{iopart}

\usepackage{graphicx}

\usepackage[T1]{fontenc}

\begin{document}

\title[An optical elevator for precise delivery of cold atoms]{An optical elevator for precise delivery of cold atoms using an acousto-optical deflector}

\author{Francesco Ferri$^{1,2}$,  Arthur La Rooij$^{1,3}$, Claire Lebouteiller$^{1}$, Pierre-Antoine Bourdel$^1$, Mohamed Baghdad$^1$, Sylvain Schwartz$^{1,4}$,  S\'{e}bastien Garcia$^{1,5}$, Jakob Reichel$^1$, Romain Long$^1$}

\address{$^1$ Laboratoire Kastler Brossel, ENS-Universit\'{e} PSL, CNRS, Sorbonne Universit\'{e}, Coll\`{e}ge de France, 24 rue Lhomond, 75005 Paris, France}
\address{$^2$ Present address: Institute for Quantum Electronics, ETH Zurich, 8093 Zurich, Switzerland}
\address{$^3$ Present address: University of Strathclyde, Department of Physics, SUPA, Glasgow G4 0NG, United Kingdom}
\address{$^4$ Present address: DPHY, ONERA, Universit\'{e} Paris-Saclay, 91123 Palaiseau, France}
\address{$^5$ Present address: Coll\`{e}ge de France, 11 Place Marcelin Berthelot, 75321 Paris, France}
\ead{long@lkb.ens.fr}


\begin{abstract}
We implement a simple method for fast and precise delivery of ultracold atoms to a microscopic device, i.e. a Fabry-Perot microcavity. By moving a single beam optical dipole trap in a direction perpendicular to the beam axis with an acousto-optical deflector, we transport up to 1 million atoms within 100$\,$ms over $1\,$cm. Under these conditions, a transport efficiency above $95\%$ is achieved with only minimal heating. The atomic cloud is accurately positioned within the microcavity and transferred into an intra-cavity optical lattice. With the addition of a secondary guiding beam, we show how residual sloshing motion along the shallow axis of the trap can be minimized.  \\ 
\\
\newpage 

\end{abstract}



\maketitle
%
%

\section{Introduction}\label{text:transport_overview}

The transport of cold neutral atoms has been realized in different experimental contexts. One common motivation is to deliver cold atoms to a ``science chamber'' benefiting from a lower vacuum pressure and/or better optical access than a ``loading chamber'' where a magneto-optical trap (MOT) is produced. Transport is also required to enable the interaction of cold atomic samples with in-vacuum microscopic devices
such as optical microcavities  \cite{Sauer2004, Nussmann2005, Khudaverdyan2008, Kim2019e, Beguin2020a, Dordevic2021a} or micromechanical resonators \cite{Hunger2010b, Gierling2011, Montoya2015}, whose size precludes the direct loading of a MOT in their vicinity. 

Different transport schemes have been implemented, which are all based on moving either magnetic of optical traps. Movable magnetic traps have been created by successively activating different pairs of coils placed along a path \cite{Greiner2001} or by mechanically moving magnetic coils on a rail \cite{Goldwin2004b}. Atom chip-based magnetic conveyors have been implemented to transport small ensemble of ultracold atoms with high positioning accuracy \cite{Hansel2001, Long2005, Gunther2005}. In all these methods, the transport setup reduces the optical access to the atoms along one or multiple directions.

The alternative method is to use optical dipole traps.  The transport of a cold atomic cloud has been achieved by moving the waist of a single-beam optical trap along the optical axis. It has been realized by translating a focusing lens \cite{Gustavson2001}, by using a pair of focus tunable lenses \cite{Leonard2014} or more recently by using a Moir\'{e} lens \cite{Unnikrishnan2021}. These methods can achieve large transport distances over tens of centimeters, but require either moving or deforming elements or achieve limited transport efficiency. Alternatively, optical conveyor belts have been developed specially in the context of Cavity Quantum Electrodynamics (CQED) experiments \cite{Kuhr2001b, Sauer2004, Nussmann2005, Khudaverdyan2008}. However, they are significantly more complex to implement, requiring costly single frequency high power lasers as well as interferometric phase stabilization, and are limited to few millimeters transport distance except for specific demanding realizations \cite{Schmid2006, Middelmann2012, Klostermann2021a}. To conclude this short overview, we note that tight optical tweezers are used for re-arrangement of single atom arrays but the typical displacements are in the 100$\,\mu$m range \cite{Beugnon2007, Lengwenus2010, Endres2016a, Barredo2016, Stuart2018b}.

For experiments requiring the interactions of cold atoms with microscopic devices, several requirements have to be fulfilled. First, the transporting trap should be tight enough for an atomic ensemble to fit inside the target device. Furthermore, the final position of the atomic cloud has to be reproducible with micrometer accuracy. In addition, one would rather avoid moving parts to prevent mechanical vibrations and consequent drifts. Finally, the transport duration between the MOT and the device should be as short as possible to limit atom losses and achieve fast experimental cycle times. For our specific experiment, the endpoint of the transport is located at 12$\,$mm from the starting point where the MOT is obtained. This target endpoint is the optical mode of a fiber Fabry-Perot microcavity (length $\simeq 140\,\mu$m, waist $\simeq 8\,\mu$m), which is a key element for the generation of multi-particle entangled states by CQED methods \cite{Haas2014, Barontini2015, McConnell2015, Welte2017}. 
  
Transporting an atomic cloud over a moderately large distance of $1\,$cm and, at the same time, position it with micrometer accuracy is challenging. To address this problem, we have developed a new and versatile solution by designing an optical elevator based on a single-beam dipole trap and an acousto-optical deflector (AOD), a device that has been used before but in different contexts \cite{Onofrio2000, Henderson2009, Rakonjac2012, Roberts2014}. This allows a robust transport with high acceleration, low loss and minimal excitation as the atomic sample is transported along a direction transverse to the beam axis, so a direction of much stronger confinement than along the beam axis. This method is simple to implement, requiring apart from the AOD only a lens and a high-power laser (not necessarily monomode). It allows us to transport a cold atomic cloud over one centimeter with acceleration up to $\sim 10\,$ m$\cdot$s$^{-2}$ and efficiency above 95$\%$, and to position it with sub-micrometer accuracy ($\sim0.6\,\mu$m). It is also robust as no moving or deforming  part is involved. 

The paper is organized as follows. In section 2, we describe the qualitative and quantitative features of our transport setup. In section 3, we characterize the setup experimentally and discuss the results. We present in section 4 an upgraded version of the setup with the addition of a guiding beam to reduce the oscillations along the shallow axis of the trap, and conclude by summarizing the results and potential further applications of our method.


\section{Implementation of the optical elevator}\label{text:optical_elevator_presentation}

The transport duration $T_{trans}$ is usually targeted to be as short a possible in order to achieve fast cycle times but different effects limit the minimal achievable duration.  The finite trap depth puts an upper limit on the maximal acceleration $a_{lim}$ of the atoms while keeping them trapped. For a given transport distance $D$, this leads to a lower limit of the transport duration $T_{lim} \propto \sqrt{\frac{D}{a_{lim}}}$. In addition, when the transport duration gets closer to the oscillation period $T_0$ of the trap,  non-adiabatic processes induce oscillations of the cloud, heating and atom loss.
If a single laser beam is used for trapping, the transport can be performed either perpendicularly or along the beam directions. We briefly analyze the different limits of the transport duration for these two geometries.

For transport along the longitudinal direction \cite{Couvert2008, Leonard2014}, the maximal acceleration scales as $a_{lim}^\| \propto \frac{U_0}{m_{Rb} z_R}$ where $U_0$ is the trap depth, $m_{Rb}$ the mass of a Rb atom, and $z_R= \frac{\pi w_0^2}{\lambda}$ is the Rayleigh range of the dipole trap beam with $w_0$ the waist and $\lambda$ the wavelength of the trapping light. We then find $\frac{T_{lim}^\| }{T_{0}} \propto \sqrt{\frac{D}{z_R}}$. A typical example of longitudinal transport can be found in Ref. \cite{Couvert2008}, where $T_{lim}^\| < T_{0} $. Non-adiabatic effects are then dominant in the longitudinal geometry and transport schemes based on shortcuts to adiabaticity can be advantageous \cite{Guery-Odelin2019}. For transport along the transverse direction, the maximal acceleration scales as $a_{lim}^\bot \propto \frac{U_0}{m_{Rb} w_0}$. One then finds $\frac{T_{lim}^\bot }{T_{0}} \propto \sqrt{\frac{D}{w_0}}$. For standard trap parameters, $T_{lim}^\bot> T_{0} $ and so the transport is limited by the finite trap depth, enabling a fast and simple transport of cold atoms, that does not required involved shortcuts to adiabaticity nor phase-stabilized standing-waves.

\begin{figure}[t]
\centering
\includegraphics[width=0.95\textwidth]{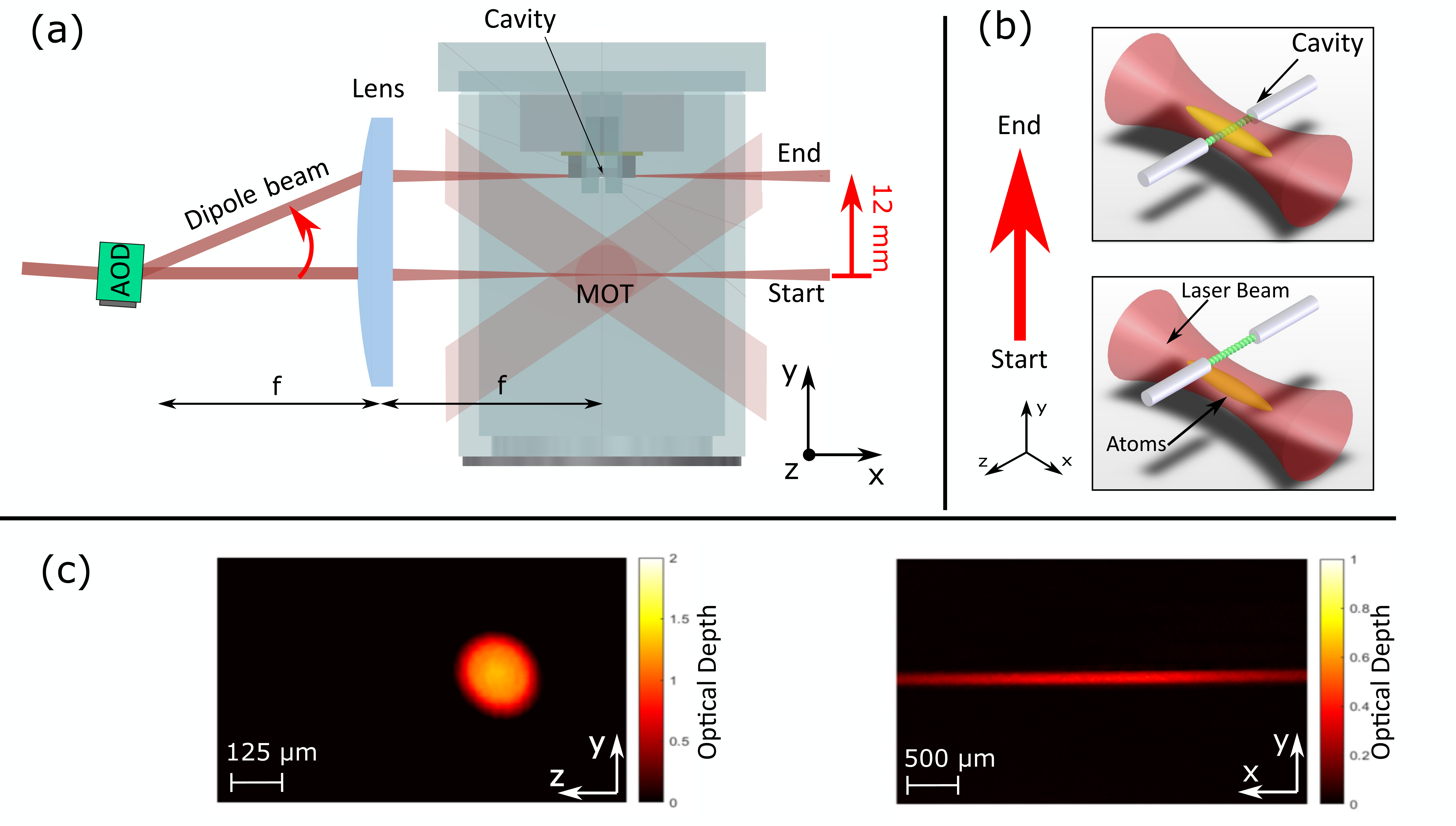}
\caption{ (a) The setup for transporting cold atoms from the MOT to the fiber mircrocavity. The method is based on a single-beam optical dipole trap at 1070$\,$nm that move due to an acousto-optical deflector (AOD). (b) Graphics showing the initial and final stage of the transport scheme. The optical trap (red) is moving the atoms (yellow) from below the cavity where the MOT is produced, up to the cavity where they can be transferred to the intra-cavity lattice (green). (c) Absorption images along two orthogonal directions of the atoms loaded in the dipole trap at the start of the transport after a time-of-flight of 0.5$\,$ms (average over 10 single-shot images).    }
\label{fig:Transport_principle}
\end{figure}

In the following, we implement a transport along the transverse direction by designing an optical elevator based on  an acousto-optical deflector (AOD) and a lens. In figure \ref{fig:Transport_principle}, the transport method is presented schematically. 
The AOD deflects a collimated red-detuned laser beam along the transverse direction: the first diffracted order is used for the optical dipole trap, while all other orders are blocked. An achromatic doublet with a focal distance $f$ focuses the collimated laser beam at the center of the science chamber along the $x$-axis. The AOD is placed at the distance $f$ behind the lens, hence sweeping the driving radio-frequency (RF) of the AOD  produces a translation of the focused beam perpendicular to the optical axis. Atoms are loaded in the dipole trap at the lower position from the magneto-optical trap and transported vertically along the  $y$-axis by 12$\,$ mm up to the microcavity.

Here we use cold $^{87}$Rubidium atoms which have a cooling transition at 780.24$\,$nm. 
The red-detuned light for the optical dipole trap is provided by a 1070$\,$nm fiber laser with a maximal output power of 50$\,$W \cite{IPG}, which is a linearly polarized multimode laser \cite{Lauber2011, Hung2015}. The laser output beam which is fiber delivered has an initial beam diameter of 5.0$\,$mm. The AOD has an aperture of 7.5$\,$mm and an efficiency above $70$\% over the whole scan angle of 85$\,$mrad \cite{AOD}.  The 2 inches diameter achromat has a focal length $f=250\,$mm, which in combination with the  scan angle of the AOD allows us to move the dipole trap by up to 14$\,$mm. The lens also sets the trap waist at $45\,\mu$m. This fulfills two important requirements: it allows the transport beam to overlap with several tens of sites of an intra-cavity optical lattice at a wavelength of $1559\,$nm; on the other hand, it is small enough to fit into the $140\,\mu$m long cavity  without depositing excessive optical power on the fibers. 

Due to the tight fit, any misalignment or extended transport duration can cause the high-power laser to damage the micromirrors. We implement a safety system to monitor both the intensity of the dipole trap beam and the frequency of the AOD (representing the transverse position of the beam). It switches off the laser beam if the dipole beam stays close the cavity for too long or when the beam intensity is too large.
The radio-frequency (RF) signals used to drive the AOD  are provided by a Direct Digital Synthesizer (DDS) \cite{DDS} controlled by an Arduino microcontroller \cite{ArduinoDue} via serial communication. The DDS is used in the linear sweep mode and the microcontroller performs a dynamical update of the DDS sweeping slope at a rate of up to 20$\,$kHz. This allows the implementation of arbitrary velocity profiles for the transport, where the effect of discretization on the atoms is fully negligible even for transport duration as short as 10$\,$ms \cite{FerriThesis}. Equivalently, discretization of the profile is negligible if the number of steps is above 100. 

\section{Transport efficiency and heating}\label{text:transport_characterization}

To load the dipole trap, the 1070$\,$nm beam is overlapped with the MOT and a 5$\,$ms molasses cooling is performed before switching off the cooling beams. Approximately $10^6$ atoms are loaded at a power of $6\,$W, which corresponds to an optical trap depth of $220\,\mu$K. The longitudinal and transverse trap frequencies are found to be $1\,$kHz and $5\,$Hz respectively. For a laser power of  6$\,$W, the temperature of the cloud after loading is about  20$\,\mu$K. Residual heating of the atoms is observed on the timescale of 1$\,$s in the static trap. However this will not play a significant role during the transport, as we achieve transport durations in the 100$\,$ms range. 

With the RF-control system described in the previous paragraph, an arbitrary velocity profile can be implemented for the transport with a high level of control.  We typically use transport ramps which are discretized in 200 time steps. Different velocity profiles have been investigated: the Blackman-Harris function \cite{Harris1978, Couvert2008}, a polynomial of degree 2, a linear function and a constant velocity profile. We observe a clear degradation of the transport if a constant velocity profile is used, while only minor differences exist between all the other profiles. In the following, we use the Blackman-Harris profile (see Appendix), which for a given transport duration leads to a slightly lowest heating of the atoms by about 2$\,\mu$K.

\begin{figure}[!ht]
    \begin{center}
        \includegraphics[width=\textwidth]{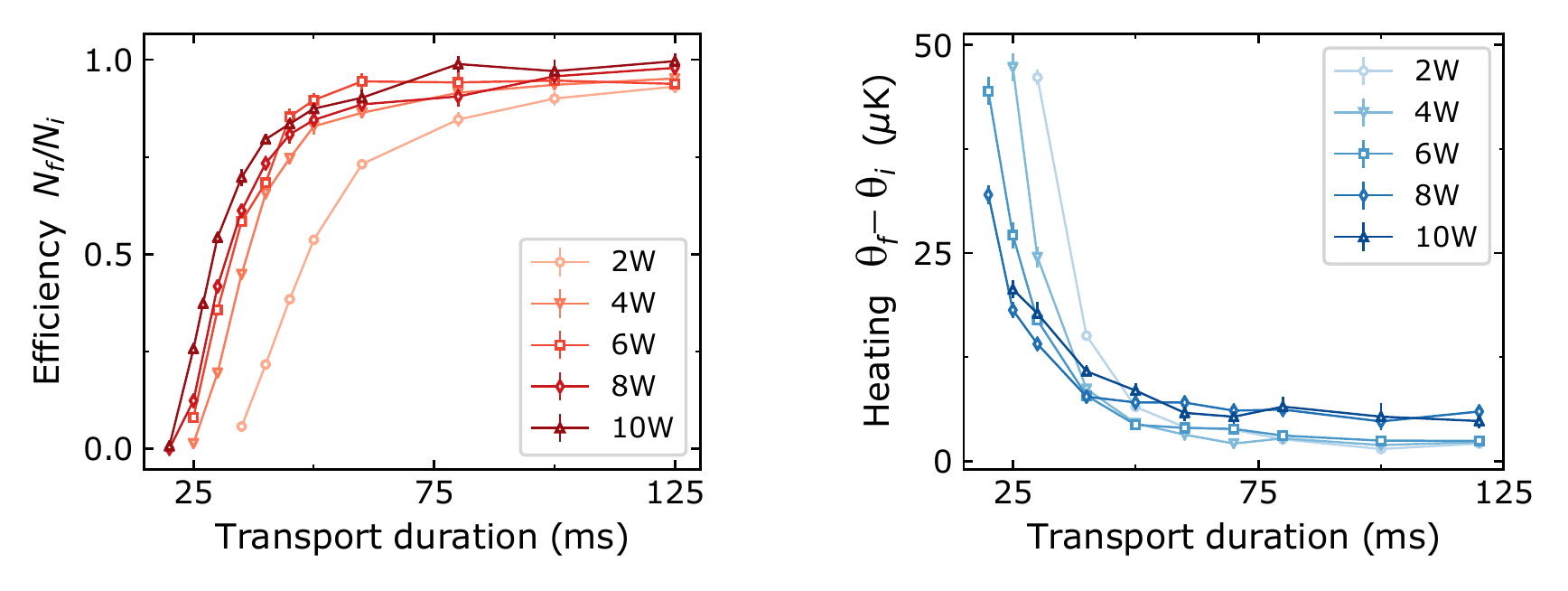}
        \label{fig:Efficiency_trans}
    \caption{\textbf{(Left)} Fraction of transported atoms as a function of the transport duration $T_{trans}$. \textbf{(Right)} Heating of the atoms $\theta_{f}-\theta_{i}$ induced by the transport.  }
    \label{fig:Transport_eff}
      \end{center}
\end{figure}

To characterize the transport with the optical elevator, the atoms are moved by 10$\,$mm and we measure the transport efficiency $\eta = \frac{N_f}{N_i}$ and the heating $\Delta \theta =\theta_{f}-\theta_{i}$ by absorption imaging. Here  $N_i$ and $N_f$ are respectively the number of atoms at the initial position and at the final position of the ramp.  The temperature increase $\Delta \theta = \theta_{f}-\theta_{i}$ of the atoms is extracted from $\theta_f$ and $\theta_i$ which are the temperature of the atoms at the end and at the start of the transport. All temperatures are measured along the $x$ axis by time-of-flight measurement. An hold time of $1\,$s was applied after the transport to allow for thermalization. In figure \ref{fig:Transport_eff}, the efficiency and the heating are shown as a function of the transport duration $T_{trans}$ for different values of the dipole laser power $P$. If $T_{trans}$ is kept larger than $100$~ms, the transport  efficiency is above $90\%$ and the heating is kept small for all laser powers. To limit the amount of heat transferred to the microcavity, a laser power of $P=6.0\,$W and a transport duration of $T_{trans}=100\,$ms are considered optimal. With these parameters,  $9.5 \times 10^5$ atoms are transported with an efficiency of $(97.0\pm0.5)\%$. The cloud temperature is increased by $(2.5\pm0.5)~\mu$K after transport. The calculated maximal acceleration is $8.6\,$m$ \cdot$s$^{-2}$, which is an order of magnitude larger than typical transport schemes along the longitudinal direction.  These results show that a fast, robust, and efficient transport can be performed over a wide range of parameters without inducing excessive heating.

As $T_{trans}$ gets shorter than $50\,$ms, losses and heating become significant. For the parameters of our transport setup, the minimal transport time $T_{lim}$ due to the finite trap depth is 4.6 times larger than the oscillation period $T_{0}$ along the transport direction. Therefore, we expect that the minimal duration for an efficient transport is limited by the finite trap depth. By computing the effective potential along the vertical direction in the reference frame of the transport beam moving against gravity for $P=6\,$W, we  find that the atoms become untrapped when moving faster than 12$\,$ms. This occurs at the time when the acceleration is maximal for a Blackman-Harris velocity profile. For transport duration 12$\,$ms $<T_{trans}<$ 40$\,$ms, the actual height of the potential barrier is strongly reduced by the acceleration in the moving frame, inducing large atom loss.

Due to the large anisotropy of the trapping potential, we also observe oscillations of the atomic cloud of $0.5\,$mm  along the longitudinal weak axis of the trap for $T_{trans}=100\,$ms (see Fig. \ref{fig:Guide_oscillations2}). We attribute the excitation of these oscillations to time-dependent thermal aberrations of the AOD and the focusing lens which affect the trapping potential during the transport. These oscillations lead to a less efficient and reproducible loading of the atoms in the cavity mode. We address this issue in the following section.

\section{Improved transport and cavity loading}\label{text:funnel}

\begin{figure}

        \includegraphics[width=\textwidth]{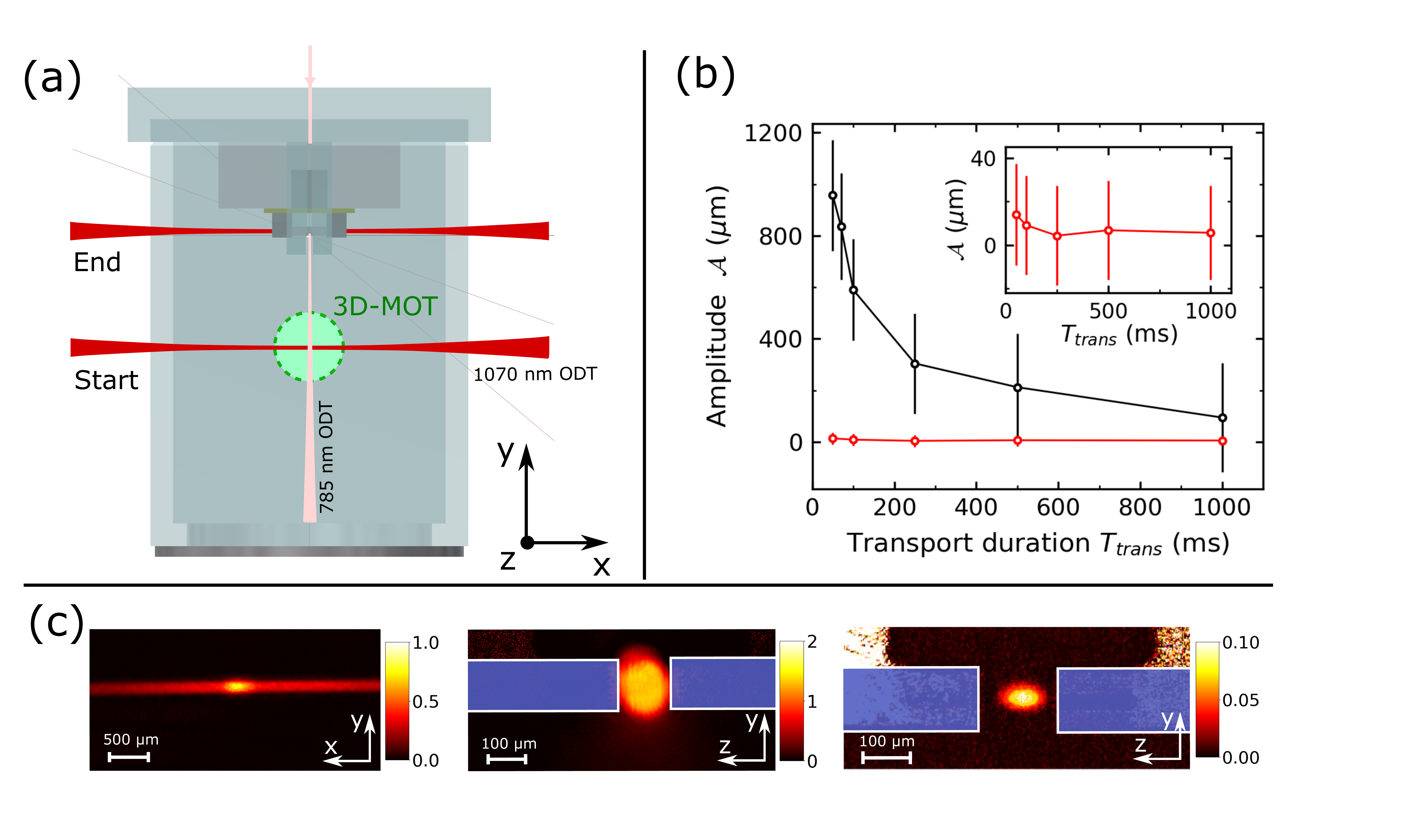}
      
    \caption{\textbf{(a)} Scheme of the additional dipole trap (wavelength $785\,$nm) for guiding the atoms in the $1070\,$nm transport trap to the fiber cavity. \textbf{(b)} Sloshing amplitude of the atoms trapped in the 1070$\,$nm trap (black, and in the crossed dipole trap (red). The atomic clouds are detected after 0.5$\,$ms free expansion and fitted with a 2D Gaussian profile in two different regions of interest; the error bars correspond to $1/4$ of the standard deviation. \textbf{(c)} Absorption images after a time-of-flight of 0.5$\,$ms of the atoms loaded in the dimple trap at the MOT position (left) and at the cavity position (middle). The right picture is an absorption image with a time-of-flight of 0.1$\,$ms of the atoms after being transferred into the 1559$\,$nm lattice . The images result from the average of 10 single-shot images. The color scales indicate the optical depth. The grey rectangles represent the two optical fibers of the microcavity.  } 
     \label{fig:Guide_oscillations2}
\end{figure}

To circumvent the aforementioned longitudinal oscillations, a second red-detuned beam is introduced, orthogonal to the transport beam, creating a crossed dipole trap. This second beam intersects both the center of the fiber cavity and the center of the $1070\,$nm dipole trap at the MOT position, see Figure \ref{fig:Guide_oscillations2} a. It acts as a guide for the sub-ensemble of atoms at the center of the elongated $1070\,$nm trap. Since the guide beam passes through the microcavity, its heating effect has to be minimized in order to protect the mirror coatings as well as to maintain the cavity stabilization. For this reason, we choose the wavelength of the guide beam to be $785.2\,$nm, which is only $5\,$nm away from the $^{87}$Rb D$_{2}$ line. As the beam waist is $50\,\mu$m,  this allows us to achieve a trap depth of 280$\,\mu$K with a reasonably low optical power of 380$\,$mW. This waist size guarantees at the same time low clipping from the cavity mirrors and a small beam divergence along the path between the cavity and the MOT.

To load the atoms in the crossed dipole trap, we proceed as previously described with the additional ramping up of the guide beam power in 25$\,$ms before the molasse phase. With this configuration and a transport duration of 100$\,$ms over 11$\,$mm,  the guide beam allows us to increase the atom number in the cavity by $30\%$. A temperature increase of the atoms by 10$\,\mu$K is also observed and deemed acceptable. It can be explained by the adiabatic compression of the cloud as the size of the guide beam is reduced from 80$\,\mu$m at the MOT to 50$\,\mu$m inside the cavity.

The main advantage of the guide beam is visible in Fig.\ref{fig:Guide_oscillations2}b, where the oscillations of the atomic ensemble along the weak axis of the 1070$\,$nm trap are shown for different transport durations. The sloshing amplitude of the atoms in the bare 1070$\,$nm trap is on the order of $0.5\,$mm for $T_{trans}=100\,$ms. In the crossed dipole trap, the position of the sub-ensemble  is stable within a few tenths of microns, which is small compared to the extension of the cloud itself, enabling the efficient and reproducible loading of atoms inside the cavity.

In our optical microcavity, the atoms are trapped in an intra-cavity standing-wave at 1559$\,$nm, which ensures an equal and maximal coupling of all the atoms with a resonant probing standing wave at 780$\,$nm \cite{Garcia2020}. To avoid irreversible damages to the cavity mirrors and to be less demanding for the cavity stabilization, we reduce the optical power of both the transport and the guide beams when moving the transport beam closer to the cavity. Starting at 2.5$\,$mm away from the cavity, we reduce exponentially their power to 15$\%$ of their initial values, which leads to an approximately isotropic trap  with calculated trap frequencies of $(400, 400, 550)\,$Hz at the cavity position. We typically end up with $10^5$ atoms in the crossed dipole trap inside the cavity at a temperature of 25$\,\mu$K.

The atoms are then transferred adiabatically from the crossed dipole trap to the intra-cavity lattice, by ramping down the transport and guide beam power while increasing the 1559$\,$nm intensity over 20$\,$ms. As the 1559$\,$nm is also used for cavity stabilization, we also have to keep a minimal amount of power inside the cavity corresponding to a trap depth of 10$\,\mu$K. In Fig. \ref{fig:Guide_oscillations2}c, we show an absorption image of a cold atoms cloud loaded in a 1$\,$mK deep intracavity lattice after a time-of-flight of 100$\,\mu$s. For such a short time, the expansion is negligible compared to the initial size.  We load typically around 1600 atoms, and the cloud size  is 60$\,\mu$m along the cavity axis, corresponding to the loading of about 80 sites with more than 10 atoms per site on average.

\section{Conclusion}

In this article, a novel scheme for transporting cold atoms into a Fabry-Perot microcavity is presented. It is based on the realization of an optical elevator with an acousto-optical deflector at the focus of a lens. We have presented a new scheme for transporting ultracold atoms that enables a fast transport at high acceleration by moving along the strong axis of the dipole trap while avoiding any mechanical vibration due to the elimination of moving parts. Clouds of $ 10^6$ atoms  are transported over a distance of $10\,$mm in $100\,$ms with efficiency above 95$\%$ and low residual heating. An additional guide beam, orthogonal to the $1070\,$nm laser, is shown to reduce the effect of atoms sloshing along the weak axis of the dipole trap. This moving crossed dipole elevator has enabled us to deliver the atomic cloud with sub-micrometer accuracy and can therefore be used to precisely position a cold atomic ensemble in or nearby microscopic devices. In our lab, this fast and highly reproducible scheme enables future studies of many-body QED entanglement generation for quantum metrology \cite{Gessner2020} and quantum simulations \cite{Davis2019}. This scheme can be used advantageously in a wide range of applications where cold atomic ensembles must be positioned with high accuracy, in particular near nanodevices such as photonic waveguides and optomechanical resonators.

\section*{Acknowledgements}
We thank Constance Poulain for careful reading of the manuscript. This project has received funding from: Agence Nationale de la Recherche (ANR) (SAROCEMA project, ANR-14-CE32-0002);
European Research Council (ERC) under the European Union's Horizon 2020 research and innovation programme Grant agreement No 671133 (EQUEMI project). This work has been supported by Region Ile-de-France in the framework of DIM SIRTEQ. S. Schwartz acknowledges funding from the European Union under the Marie Sklodowska Curie Individual Fellowship Programme H2020-MSCA-IF-2014 (project No. 658253).\\

\section*{Appendix} 

The Blackman-Harris profile of the speed used to transport the atoms over a distance $D$ in a time $T$ is given by the following equation :

The Blackman-Harris profile describes the velocity of the atoms while they move over a distance D in a time T with the following equation:
\begin{equation}\label{Blackman}
    v(t,T) =  \frac{D}{T}\left( 1-\frac{25}{21}\cos{\Big(2\pi
\frac{t}{T}\Big)} +\frac{4}{21}\cos{\Big(4\pi
\frac{t}{T}\Big)}\right).
\end{equation}

\begin{figure}[h]
\begin{center}
\includegraphics[width=0.5\textwidth]{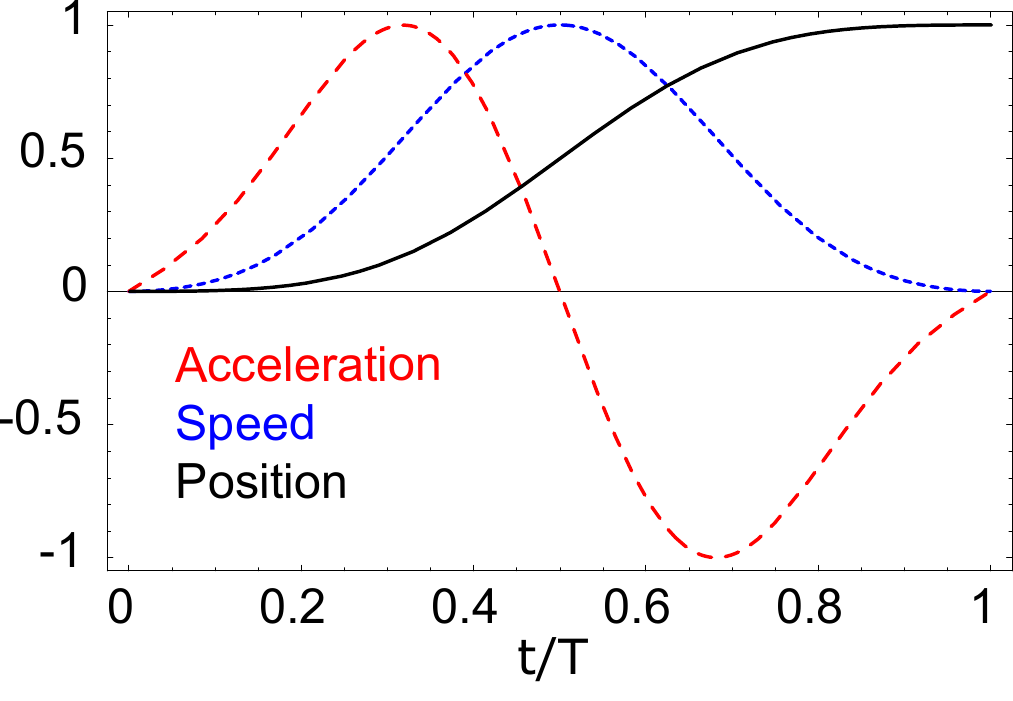}
\caption{Position profile (black line), Velocity profile scaled by $21/50$ (blue dotted line), Acceleration profile scaled by $\times 0.116$ (red dashed line) for a Blackman-Harris ramp.} \label{Blackman-fig}
\end{center}
\end{figure}

In figure \ref{Blackman-fig}, the acceleration, velocity and position are shown for $D=1$.

\newpage
\newcommand{\newblock}{}

\section*{References}

\bibliographystyle{iopart-num}
\bibliography{Atom_Transport}

\end{document}